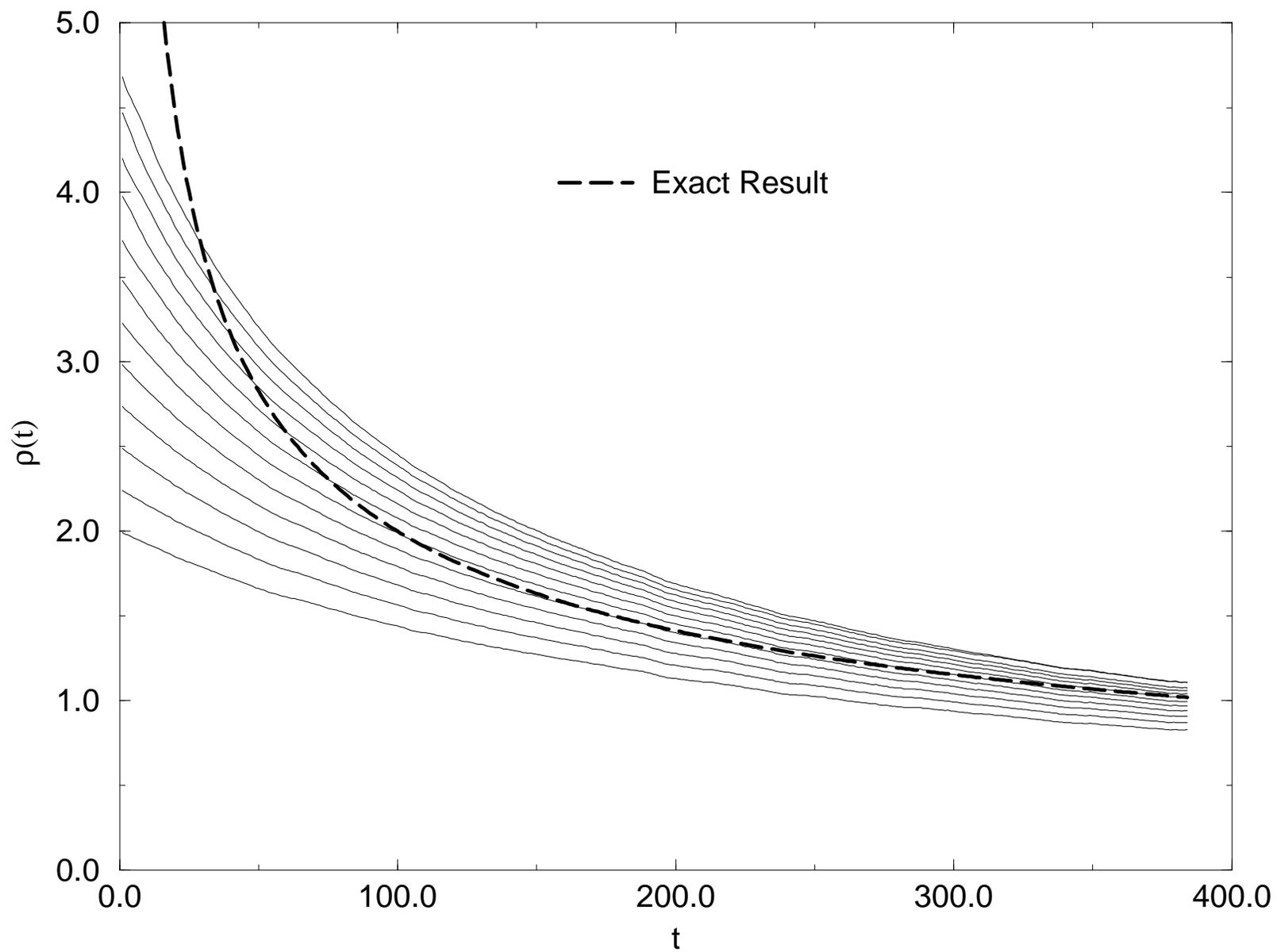
Figure I

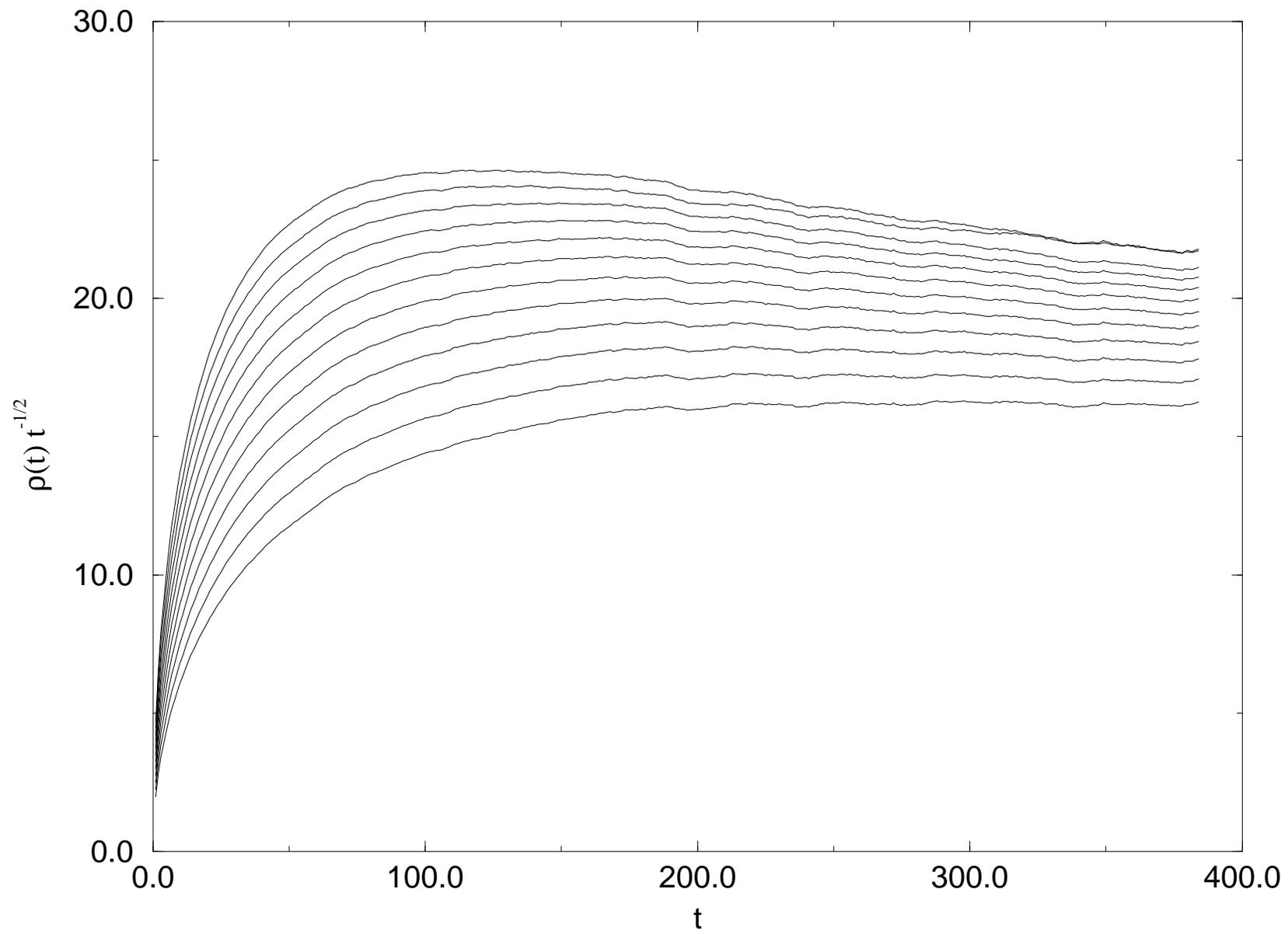

Figure II

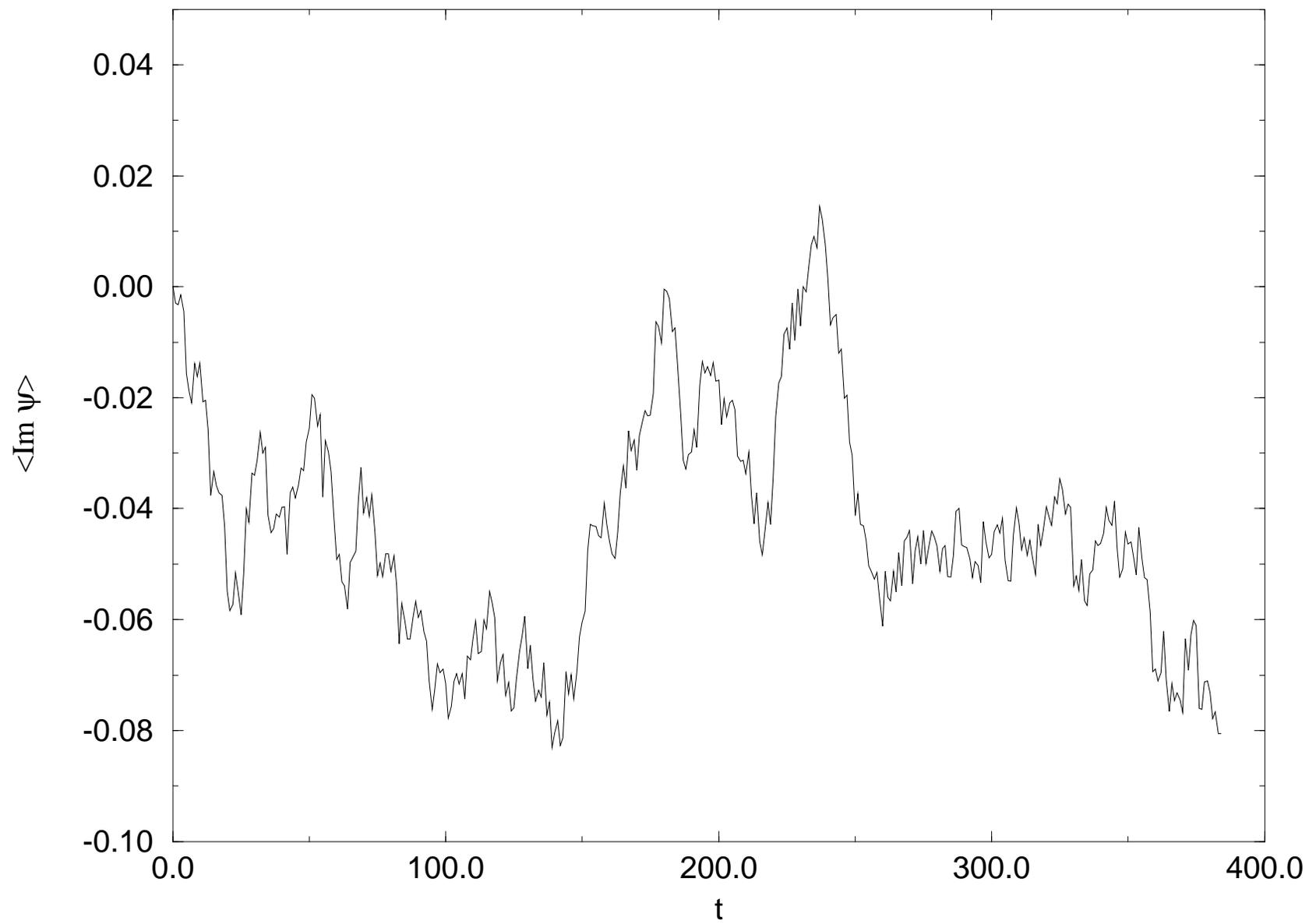

FIGURE III

# FIGURE IV

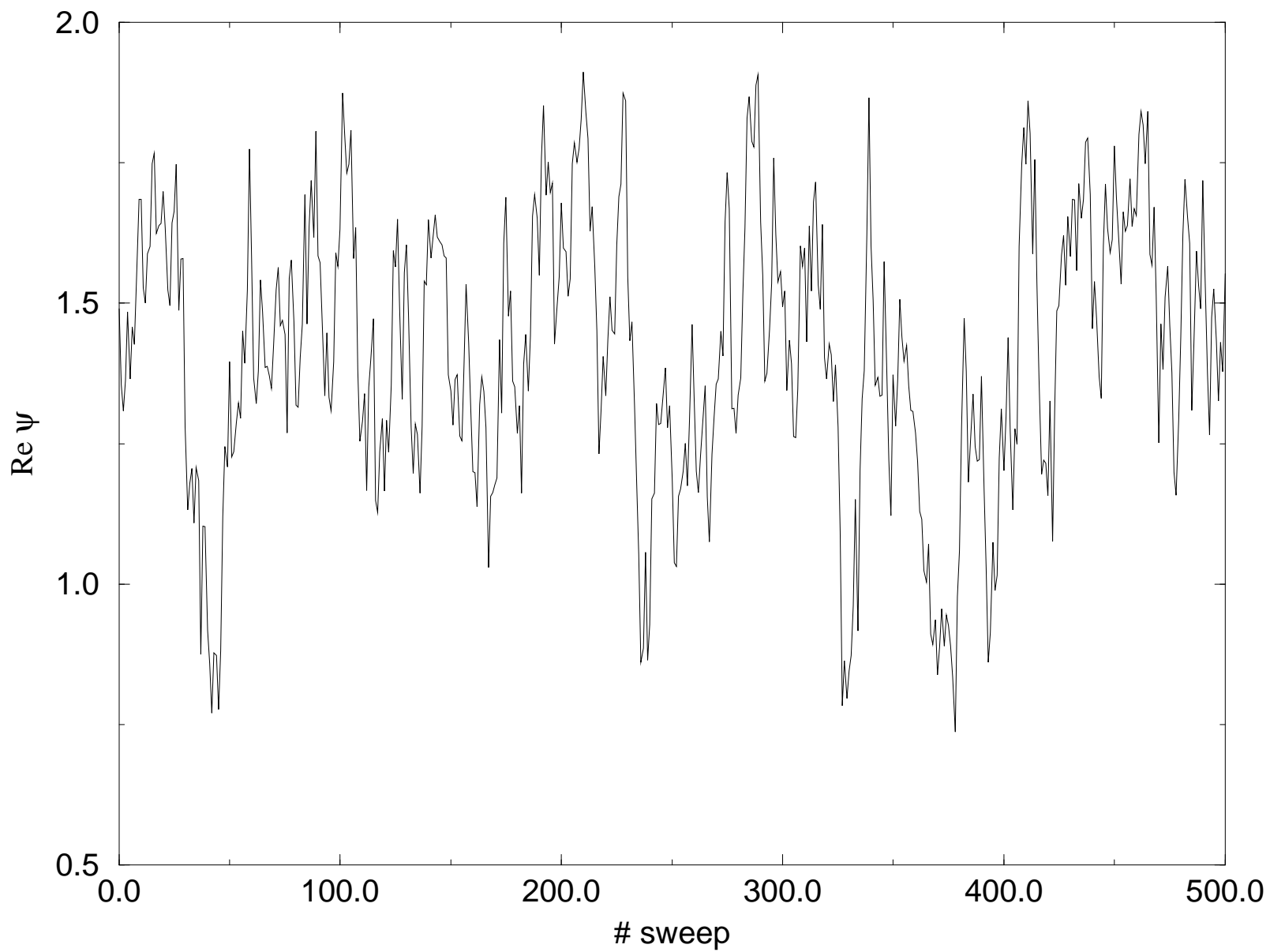

# Coherent State path-integral simulation of many particle systems

M. Beccaria[a], B. Allés[a], F. Farchioni[a]

[a] *Dipartimento di Fisica dell'Università and INFN*

*Piazza Torricelli 2, I-56100 Pisa, Italy*

## Abstract

The coherent state path integral formulation of certain many particle systems allows for their non perturbative study by the techniques of lattice field theory. In this paper we exploit this strategy by simulating the explicit example of the diffusion controlled reaction $A + A \to 0$. Our results are consistent with some renormalization group-based predictions thus clarifying the continuum limit of the action of the problem.

# I. INTRODUCTION

An approach to the non perturbative definition and study of quantum field theories is given by path integral quantization. Lattice field theory is based on such a formulation. The functional integral is built from the infinitesimal propagation of particles among states of a definite basis. If the Hamiltonian is given in terms of annihilation and creation operators, then the most natural (overcomplete) basis is that of coherent states [1,2].

A relevant example is that of the so called diffusion-controlled chemical reactions [3]. These are physical processes describing $N$ particle species $A_1$, $A_2$, $\cdots$ diffusing on a lattice and undergoing annihilation-creation reactions of the form

$$n_1 A_1 + \cdots n_N A_N \to m_1 A_1 + \cdots m_N A_N. \tag{1.1}$$

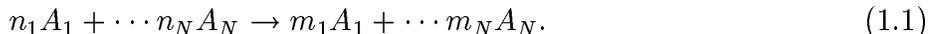

The configuration space of this system has a structure resembling that of the Fock space of relativistic particles. The time evolution of the probability distribution of the particles is described by a Master Equation and the evolution operator is built from a non hermitian Fokker-Planck *hamiltonian* written in terms of creation-annihilation operators. Statistical averages are traded in a standard way for quantum expectation values [4] and the (non unitary) evolution may be explicitely solved by a coherent state path-integral [5].

Renormalization Group techniques can be used: this approach has been applied successfully to the *formal* continuum limit of several models, a typical prediction being the behaviour of the particle densities as a function of time [6–9].

However, the comparison with numerical data is often non trivial because numerical simulations are performed under conditions slightly but significantly different than those of the analytic computations. Typical examples may be an infinite reaction rate or a limited single site occupancy (see [10] for a study of the finite rate effects).

An interesting alternative to the direct microscopic simulation is to study the coherent state formulation by the usual tools of lattice field theory. This allows for a direct verification of the renormalized perturbation theory results.

This strategy faces several drawbacks. First the action in the path integral is complex as the Fokker-Planck hamiltonian is not hermitian. The convergence properties of simulation algorithms for complex actions in interacting models are in general not guaranteed [11]. On the other hand, the continuum limit of the discrete model presents some ambiguities which may be seen as operator ordering. It is not clear a priori whether these ambiguities can modify the resulting measurable quantities.

The aim of this paper is two-fold. First we shall analyze analytically the behaviour of an exactly solvable model: the free coherent state path integral from the point of view of its numerical simulation. Secondly we shall perform an explicit simulation on a non-trivial model, the reaction $A + A \to 0$, in order to verify the relevance of the above-mentioned problems.

In section II we shall review the coherent state path-integral formulation of a problem defined by a hamiltonian in terms of creation and annihilation operators. We will introduce the ambiguity in the continuum limit and will show that two actions (identical in that limit but different in the discrete version of the theory) display a rather opposite behaviour under the Langevin algorithm during the simulation. In section III we will introduce the $A+A \to 0$ problem and the numerical simulation together with its results.

## II. COHERENT STATE PATH-INTEGRAL

Let us consider a one-dimensional quantum harmonic oscillator with unit pulsation and hamiltonian

$$\hat{H} = \hat{a}^\dagger \hat{a} + \frac{1}{2}, \tag{2.1}$$

where $\hat{a}^\dagger$ and $\hat{a}$ are creation and annihilation operators satisfying the canonical commutation relation

$$[\hat{a}, \hat{a}^\dagger] = 1. \tag{2.2}$$

Coherent states $|z\rangle$ are defined as eigenvectors of the destruction operator

$$|z\rangle = \exp\left(-|z|^2/2 + z\hat{a}^\dagger\right)|0\rangle, \qquad (2.3)$$

$$\hat{a}|z\rangle = z|z\rangle, \qquad (2.4)$$

where $|0\rangle$ is the vacuum. With this normalization we have

$$\langle w|z\rangle = \exp\left(\bar{w}z - |z|^2/2 - |w|^2/2\right), \qquad (2.5)$$

$$1 = \int \frac{d^2z}{\pi}|z\rangle\langle z|. \qquad (2.6)$$

The Euclidean propagator for an arbitrary hamiltonian $\hat{H}(\hat{a}, \hat{a}^\dagger)$ is

$$U(z'', t|z', 0) = \langle z''|\exp(-t\hat{H})|z'\rangle, \qquad (2.7)$$

and its expansion when $t \to 0$ can be used to give a lattice path integral definition of $U$

$$U^{(N)}(z'', t|z', 0) = \qquad (2.8)$$

$$= \int \frac{d^2z_1 \cdots d^2z_N}{\pi^N} \exp \sum_{n=0}^{N} \left\{\frac{1}{2}[(\bar{z}_{n+1} - \bar{z}_n)z_n - \bar{z}_{n+1}(z_{n+1} - z_n)] - \epsilon H(\bar{z}_{n+1}, z_n)\right\}, \qquad (2.9)$$

where

$$z_0 = z', \quad z_{N+1} = z'', \qquad (2.10)$$

$$\epsilon(N+1) = t, \qquad (2.11)$$

$$H(\bar{w}, z) = \langle w|H|z\rangle/\langle w|z\rangle. \qquad (2.12)$$

The limit

$$\lim_{N \to \infty} U^{(N)} = U, \qquad (2.13)$$

is justified in terms of Trotter's formula just as in the usual coordinate basis path integral. The formal continuum limit of the above functional integral is often written

$$U = \int \mathcal{D}z\mathcal{D}\bar{z}\, e^{-S}, \quad S = \int dt \left\{\frac{1}{2}[-\dot{\bar{z}}z + \bar{z}\dot{z}] + H(\bar{z}, z)\right\}, \qquad (2.14)$$

and it is used as a starting point for subsequent analysis, e.g. perturbation expansion. However it must be kept in mind that the above expression stands for the lattice action

$$S^{(N)} = \sum_{n=0}^{N} \left\{ \frac{1}{2} \left[ -(\bar{z}_{n+1} - \bar{z}_n) z_n + \bar{z}_{n+1}(z_{n+1} - z_n) \right] + \epsilon H(\bar{z}_{n+1}, z_n) \right\}, \tag{2.15}$$

and not for the naive discretization

$$\tilde{S}^{(N)} = \sum_{n=0}^{N} \left\{ \frac{1}{2} \left[ -(\bar{z}_{n+1} - \bar{z}_n) z_n + \bar{z}_n(z_{n+1} - z_n) \right] + \epsilon H(\bar{z}_{n+1}, z_n) \right\}, \tag{2.16}$$

the difference being the lattice operator

$$\delta S^{(N)} = \frac{1}{2} \sum_{n=0}^{N} |z_{n+1} - z_n|^2. \tag{2.17}$$

The relevance of the above term has already pointed out in [12] in the study of the harmonic oscillator and the trace

$$\text{Tr}\left(e^{-\beta \hat{H}}\right) = \int \frac{d^2 z}{\pi} \langle z | e^{-\beta \hat{H}} | z \rangle, \tag{2.18}$$

which is associated to the path integral with periodic boundary conditions. In this paper we shall be concerned with the Feynman propagator $U(z'', t | z', 0)$ with fixed boundary conditions. The initial state $|z'\rangle$ contains all the information about the initial set up of the diffusive system we want to study. The final state $|z''\rangle$ is somewhat more arbitrary. The effect of the different discretizations will be examined by computing $U$ and also a relevant two point function of the $\hat{a}$ and $\hat{a}^\dagger$ operators. Of course, the interest in $\tilde{S}$ is purely mathematical since that form of the action has no physical relevance.

Apart from the subtleties associated to the discretization, there is another difficulty. In realistic applications, both the above actions are complex and their non perturbative (numerical) study is difficult. A possible approach to their Monte Carlo simulation relies on the Langevin algorithm. In the following subsections we shall show for the free theory that action $S^{(N)}$ is stable under this algorithm and can give sensible results; on the other hand, a simulation with the action $\tilde{S}^{(N)}$ would be unstable.

### A. Structure of the Action

Apart from additive constants, the action of the harmonic oscillator is

$$S^{(N)} = \sum_{n=0}^{N} \left\{ \frac{1}{2}|z_{n+1}|^2 + \frac{1}{2}|z_n|^2 - (1-\epsilon)\bar{z}_{n+1}z_n \right\}. \tag{2.19}$$

On the other hand, the modified action is

$$\tilde{S}^{(N)} = \sum_{n=0}^{N-1} \left\{ \frac{1}{2}\bar{z}_n z_{n+1} + (\epsilon - 1/2)\bar{z}_{n+1}z_n \right\}, \tag{2.20}$$

We could introduce real fields suitable for the simulation, but for analytical computations we prefer to work with the $z$ and $\bar{z}$ variables and consider apart from constants

$$S = \bar{z}^T A z + \bar{C}^T z + \bar{z}^T B, \tag{2.21}$$

Let us give the expression of $A$, $A^{-1}$, $B$ and $\bar{C}$ for the two actions $S$ and $\tilde{S}$. For the action $S$ we have

$$A = \begin{pmatrix} 1 & & & 0 \\ -\theta & 1 & & \\ & -\theta & 1 & \\ 0 & & \ddots & \ddots \end{pmatrix}, A^{-1} = \begin{pmatrix} 1 & & & 0 \\ \theta & 1 & & \\ \theta^2 & \theta & 1 & \\ & \cdots & & \end{pmatrix}, B = \begin{pmatrix} -\theta z' \\ 0 \\ \vdots \end{pmatrix}, \bar{C} = \begin{pmatrix} \vdots \\ 0 \\ -\theta \bar{z}'' \end{pmatrix}. \tag{2.22}$$

where $\theta = 1 - \epsilon$. For the action $\tilde{S}$ we have

$$A = \begin{pmatrix} 0 & \alpha & & 0 \\ \beta & 0 & \alpha & \\ & \beta & 0 & \ddots \\ 0 & & \ddots & \ddots \end{pmatrix}, B = \begin{pmatrix} \beta z' \\ 0 \\ \vdots \\ 0 \\ \alpha z'' \end{pmatrix}, \bar{C} = \begin{pmatrix} \alpha \bar{z}' \\ 0 \\ \vdots \\ 0 \\ \beta \bar{z}'' \end{pmatrix}, \tag{2.23}$$

where $\alpha = 1/2$ and $\beta = \epsilon - 1/2$. The inverse matrix exists only for even $N$ and is given by the formula

$$A^{-1}_{mn} = \begin{cases} 0 & n - m \text{ even} \\ (-1)^{\frac{n-m-1}{2}} \frac{1}{\beta} \left(\frac{\alpha}{\beta}\right)^{\frac{n-m-1}{2}} & n - m > 0 \\ (-1)^{\frac{m-n-1}{2}} \frac{1}{\alpha} \left(\frac{\beta}{\alpha}\right)^{\frac{m-n-1}{2}} & n - m < 0 \end{cases} \tag{2.24}$$

## B. Spectrum and Langevin Simulation

The Langevin algorithm for a lattice field theory is a way of generating field configurations distributed according to the discrete measure

$$\mathcal{D}\phi = e^{-S(\phi_1,\cdots,\phi_N)} \prod_{n=1}^{N} d\phi_n, \qquad (2.25)$$

where $\phi_n$ are the discrete real degrees of freedom in the lattice approximation. If we consider the flat case (so $d\phi$ is the flat Lebesgue measure) the algorithm introduces a fictitious time $\tau$ and evolves the configurations $\phi^{(\tau)}$ according to the Markov chain

$$\phi_k^{(\tau+\Delta\tau)} = \phi_k^{(\tau)} - \Delta\tau \frac{\partial S}{\partial \phi_k}(\phi^{(\tau)}) + \sqrt{2\Delta\tau}\xi_k^{(\tau)}, \qquad (2.26)$$

where $\xi^{(\tau)}$ is a white gaussian noise with two point correlation matrix

$$\langle \xi_k^{(\tau)} \xi_{k'}^{(\tau')} \rangle = \delta_{kk'}\delta_{\tau\tau'}. \qquad (2.27)$$

These configurations tend to be distributed according to the above weight in the limit $\Delta\tau \to 0$. If the fields are real, but the action is complex, we can still perform the algorithm updates by complexifying the field (but not the noise). The conditions under which this scheme gives correct results for an interacting theory with complex action are not known in general (see [11] for a mathematical discussion and an explicit application to the quantized Hall effect).

To start with, let us check when the free action is correctly simulated. We will see that even in this trivial case the previous algorithm works for the action $S$ and not for the $\tilde{S}$. For the above quadratic action the following statement holds: the n-point correlation function $\langle \phi(t_1)\phi(t_2)\cdots \rangle$ converge to their proper value in the limit $\Delta\tau \to 0$ if and only if the spectrum of $1 - \Delta\tau A$ is strictly inside the unit circle in this limit. To illustrate this statement let us show the result of a Langevin simulation on the 1-point function whose continuum value vanishes. The Langevin equation is (we use an integer number to label the discrete fictitious time)

$$\langle \phi^{(n+1)} \rangle = M \langle \phi^{(n)} \rangle, \quad M = 1 - \Delta\tau A. \tag{2.28}$$

Hence

$$\langle \phi^{(n)} \rangle = M^n \langle \phi^{(0)} \rangle, \tag{2.29}$$

On the other hand, if the maximum modulus of the set of eigenvalues of $M$ is less than 1 then $M^n v \to 0$ as $n \to \infty$. This follows from the fact that $M$ is always similar to the direct sum of Jordan blocks associated to the eigenvalues $\lambda$ and of the form

$$I(\lambda) = \begin{pmatrix} \lambda & 1 & 0 & \ldots \\ 0 & \lambda & 1 & \ldots \\ 0 & 0 & \lambda & \ldots \end{pmatrix}, \tag{2.30}$$

and it is easy to see that $I(\lambda)^n \to 0$ if $n \to \infty$ and $|\lambda| < 1$.

Let us examine the spectral structure of the actions $S$ and $\tilde{S}$. In the case of $S$ it is straightforward to show that

$$\det(A + \gamma) = (1 + \gamma)^N, \tag{2.31}$$

which implies that the spectrum of $1 - \Delta\tau A$ is the single point

$$\lambda = 1 - \Delta\tau. \tag{2.32}$$

This result in turn implies stability of the Langevin algorithm according to the above remarks. The analogous study for the action $\tilde{S}^{(N)}$ is more complicated. The determinant

$$p_N(\gamma) = \det(\gamma + A), \tag{2.33}$$

satisfies

$$p_N(\gamma) = \gamma p_{N-1}(\gamma) - \alpha\beta p_{N-2}(\gamma), \tag{2.34}$$

$$p_0 = 1, \tag{2.35}$$

$$p_1 = \gamma. \tag{2.36}$$

The solution is

$$p_N(\gamma) = \frac{1}{\alpha\beta - \mu^2}\left(-\mu^{N+2} + \frac{(\alpha\beta)^{N+1}}{\mu^N}\right), \quad \mu = \frac{\gamma + \sqrt{\gamma^2 - 4\alpha\beta}}{2}, \qquad (2.37)$$

and the zeroes of $p_N(\gamma)$ are given by the equation

$$p_N(\gamma) = 0 \Rightarrow \left(\frac{\alpha\beta}{\mu^2}\right)^{N+1} = 1. \qquad (2.38)$$

Notice that if $\gamma$ is a solution, so is $-\gamma$. The eigenvalues of $1 - \Delta\tau A$ may be written in the form $\lambda = 1 - \Delta\tau\gamma$ where $\gamma$ are determined by the equation $p_N(\gamma) = 0$. All the non-zero roots of this equation appear in doublets $\pm\gamma$. This means that the spectrum of $1 - \Delta\tau A$ cannot be strictly inside the unit circle.

In the appendix, we compute the Feynman propagator and a two point function by using the two different actions showing further problems in the physical meaning of the action $\tilde{S}$.

## III. DIFFUSION CONTROLLED CHEMICAL REACTIONS

### A. Field Theoretical Formulation

Let us now turn to an explicit non trivial example in order to show that the direct simulation of the coherent state path integral is feasible. We have considered one of the diffusion-controlled chemical reactions of [7]. It describes equal particles $A$ diffusing isotropically in one dimension and interacting by means of the reaction

$$A + A \to 0. \qquad (3.1)$$

Mean field theory does not apply for dimension $d \leq 2$ and fluctuations are very relevant.

Let us briefly review how the coherent state path integral arises in the treatment of this problem. This procedure is by now standard and we recall it in a few lines. Let $P(\{n\})$ be the probability distribution of the particle configuration $\{n\}$. The notation is $\{n\} = (n_1, \cdots, n_L)$ for a lattice with side $L$. Let us set to unity the spatial lattice spacing; the evolution of $P$ is described by the Master Equation

$$\frac{\partial}{\partial t}P(\{n\},t) = \hat{\Omega}P(\{n\},t), \tag{3.2}$$

where the operator $\hat{\Omega}$ is

$$\hat{\Omega} = \mathcal{D}\sum_{i,j}[(n_j+1)\hat{T}_i^{-1}\hat{T}_j - n_i] + \lambda\sum_i[(n_i+2)(n_i+1)\hat{T}_i^2 - n_i(n_i-1)]; \tag{3.3}$$

In this equation $\mathcal{D}$ is the diffusion constant and $\lambda$ is the annihilation rate constant. The sum in $j$ runs over the neighbours[1] of the site $i$ and the shift operator $\hat{T}$ is defined by

$$\hat{T}_i^k P(\{n\},t) = P(n_1, n_2, \cdots, n_{i-1}, n_i + k, n_{i+1}, \cdots, t). \tag{3.4}$$

To each site we associate a quantum harmonic oscillator with its creation-annihilation operators $\hat{a}_i$ and $\hat{a}_i^\dagger$. We then introduce the state

$$|\phi(t)\rangle = \sum_{\{n\}} P(\{n\},t) \prod_i (\hat{a}_i^\dagger)^{n_i}|0\rangle. \tag{3.5}$$

We can call such a state a probabilistic state in order to emphasize the property

$$\sum_{\{n\}} P(\{n\},t) = 1, \tag{3.6}$$

The time evolution of $|\phi\rangle$ is governed by the Schrodinger equation

$$-\frac{\partial}{\partial t}|\phi(t)\rangle = \hat{H}|\phi(t)\rangle, \tag{3.7}$$

with Hamiltonian

$$\hat{H} = -\mathcal{D}\sum_{i,j}\hat{a}_i^\dagger(\hat{a}_j - \hat{a}_i) - \lambda\sum_i(1 - (\hat{a}_i^\dagger)^2)\hat{a}_i^2. \tag{3.8}$$

Finally, one introduces the so called projection state

$$\langle\Pi| = \langle 0|\prod_i e^{\hat{a}_i}, \tag{3.9}$$

---

[1] we shall be concerned with hypercubic lattices where the neighbours of a site $P$ are the sites at distance from $P$ equal to the lattice spacing.

such that the statistical averages satisfy

$$\sum_{\{n\}} P(\{n\}, t) F(\{n\}) = \langle \Pi | \hat{F} e^{-t\hat{H}} | \phi(0) \rangle. \tag{3.10}$$

Given $F(\{n\})$, the explicit form of $\hat{F}$ is obtained substituting $n$ by $\hat{a}^\dagger \hat{a}$. Moreover, if $\hat{F}$ is normal ordered, then the creation operators may be skipped because

$$\langle \Pi | \hat{a}^\dagger = \langle 0 | e^{\hat{a}} \hat{a}^\dagger e^{-\hat{a}} e^{\hat{a}} = \langle 0 | (\hat{a}^\dagger + [\hat{a}, \hat{a}^\dagger]) e^{\hat{a}} = \langle \Pi |. \tag{3.11}$$

For instance, the density operator is just the operator

$$\hat{\rho} = \frac{1}{L} \sum_n \hat{a}_n. \tag{3.12}$$

Let us remark that for any probabilistic $|\phi\rangle$ we have

$$\langle \Pi | e^{-t\hat{H}} | \phi \rangle = 1, \tag{3.13}$$

the probability states form an overcomplete basis of the state space, hence we obtain the important probability conservation relation

$$\langle \Pi | \hat{H} = 0. \tag{3.14}$$

Our goal is to determine the anomalous exponent $\gamma$ of the density of particles $\rho(t)$. If $\mathcal{D}$ is the diffusion constant, the theoretical prediction for the density in 1 dimension and in the $t \to \infty$ limit is [13]

$$\lim_{t \to +\infty} [(\mathcal{D}t)^\gamma \rho(t)] = A, \quad A = \frac{1}{\sqrt{8\pi}}, \quad \gamma = \frac{1}{2}. \tag{3.15}$$

We consider an initial state such that the occupance probability distribution at each site is Poissonian with average occupation number $\bar{n}$. The initial state is thus proportional to a coherent state with $z = \bar{n}$ since

$$e^{-\bar{n}} \sum_k \frac{\bar{n}^k}{k!} (\hat{a}^\dagger)^k |0\rangle = e^{-\bar{n} + \hat{a}^\dagger \bar{n}} |0\rangle = e^{-\bar{n} + \bar{n}^2/2} |\bar{n}\rangle. \tag{3.16}$$

We can write

$$\rho(t) = \langle\Pi|\hat{\rho}\exp(-t\hat{H})|\bar{n}\rangle e^{-\bar{n}+\bar{n}^2/2} = \frac{\langle\Pi|\exp(-(t_f-t)\hat{H})\,\hat{\rho}\exp(-t\hat{H})|\bar{n}\rangle}{\langle\Pi|\exp(-t_f\hat{H})|\bar{n}\rangle}, \qquad (3.17)$$

The above quantity may be computed non perturbatively by a Monte Carlo simulation on a lattice with temporal extension $t_f$ and by measuring at each update the value of $\rho$ as a function of time. The evolution in time up to $t_f$ may be included precisely because of result (3.14).

### B. Numerical Simulation

We made use of the action $S$ in order to perform the Monte Carlo simulation. The integration variables were called $\psi$ and $\bar{\psi}$. We used a rectangular lattice with spatial and temporal sizes $L$ and $T$ respectively. The complex action is

$$S[\psi,\bar{\psi},L,T] = \sum_{x=1}^{L}\left\{\sum_{t=1}^{T}\left[\frac{1}{2}\bar{\psi}_{t+1,x}(\psi_{t+1,x}-\psi_{t,x}) - \frac{1}{2}\psi_{t,x}(\bar{\psi}_{t+1,x}-\bar{\psi}_{t,x})+\right.\right. \qquad (3.18)$$

$$\left.\left. -\epsilon\mathcal{D}\bar{\psi}_{t+1,x}\hat{\nabla}_x^2\psi_{t,x} - \epsilon\lambda(1-\bar{\psi}_{t+1,x}^2)\psi_{t,x}^2\right] - \psi_{N,x}\right\}. \qquad (3.19)$$

The term $\hat{\nabla}_x^2\psi_{t,x}$ is the finite difference $\psi_{t,x+1} - 2\psi_{t,x} + \psi_{t,x-1}$. The asymptotic state is the vacuum which we put at time $T+1$; the projection state is at time $N$. We have said that the action is complex: this means that the imaginary unit does not cancel if we define $q$ and $p$ (we omit the $(t,x)$ label) by

$$\psi = q + ip, \quad \bar{\psi} = q - ip, \qquad (3.20)$$

and write the action as a function of $(q,p)$. The Langevin equations are

$$\frac{\partial q}{\partial \tau} = -\frac{\partial S}{\partial q} + \xi^{(q)}, \qquad (3.21)$$

$$\frac{\partial p}{\partial \tau} = -\frac{\partial S}{\partial p} + \xi^{(p)}, \qquad (3.22)$$

with independent noises $\xi^{(q)}$ and $\xi^{(p)}$. Since the action is complex, the variables $(q,p)$ may wander in the complex plane. In terms of the (also) complex variables $(\psi,\bar{\psi})$ we have the Langevin equations

$$\frac{\partial \psi}{\partial \tau} = -2\frac{\partial S}{\partial \bar{\psi}} + \xi^{(q)} + i\xi^{(p)}, \qquad (3.23)$$

$$\frac{\partial \bar{\psi}}{\partial \tau} = -2\frac{\partial S}{\partial \psi} + \xi^{(q)} - i\xi^{(p)}, \qquad (3.24)$$

The discrete form of these equations describe the update of the configuration from the Langevin time $n$ to the time $n+1$. They are

$$\psi_{t,x}^{(n+1)} = \psi_{t,x}^{(n)} + 2\Delta\tau \bar{F}_{t,x}(\psi^{(n)}, \bar{\psi}^{(n)}) + \sqrt{2\Delta\tau}(\xi_{t,x}^{(q)} + i\xi_{t,x}^{(p)}), \qquad (3.25)$$

$$\bar{\psi}_{t,x}^{(n+1)} = \bar{\psi}_{t,x}^{(n)} + 2\Delta\tau F_{t,x}(\psi^{(n)}, \bar{\psi}^{(n)}) + \sqrt{2\Delta\tau}(\xi_{t,x}^{(q)} - i\xi_{t,x}^{(p)}), \qquad (3.26)$$

$$(3.27)$$

where

$$F_{t,x} = \bar{\psi}_{t,x} - \bar{\psi}_{t+1,x} + \epsilon\mathcal{D}(2\bar{\psi}_{t+1,x} - \bar{\psi}_{t+1,x-1} - \bar{\psi}_{t+1,x+1}) - 2\epsilon\lambda\psi_{t,x}(1 - \bar{\psi}_{t+1,x}^2) - \delta_{t,N}, \qquad (3.28)$$

$$\bar{F}_{t,x} = \psi_{t,x} - \psi_{t-1,x} + \epsilon\mathcal{D}(2\psi_{t-1,x} - \psi_{t-1,x-1} - \psi_{t-1,x+1}) + 2\epsilon\lambda\bar{\psi}_{t,x}\psi_{t-1,x}^2. \qquad (3.29)$$

We remark the very important fact that $\bar{F} \neq F^*$. We assume periodic boundary conditions in the spatial direction. On the temporal boundary $t = T$ we have set $\bar{\psi}$ to zero. At $t = 0$ we assume the above mentioned coherent state $|\bar{n}\rangle$.

As a minor point, we would like to stress a well known property of the Langevin algorithm. The only required mathematical properties of the gaussian noise $\xi$ are its first two moments. Therefore, instead of a normal gaussian random number we can use $\sqrt{3}(2u - 1)$ where $u$ is a random number with flat distribution in $[0, 1]$. This trick makes the simulation pretty faster.

Moreover, concerning the issue of numerical stability, we remark that some runs were performed by using single precision arithmetics, showing no apparent discrepancy with the double precision results.

## IV. RESULTS AND DISCUSSION

For the numerical simulation we have considered a lattice with $L \times T = 80 \times 384$. We chose a non linear coupling $\lambda = 0.001$, a physical time step $\epsilon = 0.01$ and a dimensionless diffusion constant $\mathcal{D} = 0.01$.

We ran the simulation starting from several values of the initial density to check that the asymptotic density amplitude is independent of the initial density. We performed about $10^7$ updates with the Langevin time step $\Delta \tau = 0.0025$. Measurements were separated by $10^4$ decorrelation sweeps.

Concerning the statistical errors, we found that the intrinsic standard deviation of $\rho(t)$ (the raw Monte Carlo fluctuations of data) increases roughly like $t^{1/4}$ [2]. The standard deviation must be corrected with the residual autocorrelation of the samples which we found to depend very little on $t$. The resulting total errors are very small on the scale of the following figures and they are not shown. We also remark that the values of $\rho(t)$ corresponding to different $t$ are also rather uncorrelated because of self averaging being the results of spatial averages.

In Figure (I), we collect the time behaviour of the density for different initial densities. The fact that error bars are very small can be seen by the rather small fluctuations of the curves.

In Figure (II) we show the behaviour of the effective amplitude $\rho(t)\sqrt{t}$. The imaginary part of the density was always negligible.

In Figure (III) we show as an example the average value of $\text{Im}\psi(t)$ and finally in Figure (IV) we show a portion of a typical time history from which we extracted $\rho(t)$.

Data reproduce well the exact $-1/2$ exponent. Concerning the amplitude $A$, the theoretical universal amplitude is 19.95 in our units whereas in the figure the various curves seem to settle around 18. This value is chosen as follows: we see that the density profiles in Figure (II) show two different qualitative behaviours, some of them are monotonically increasing with time, whereas the others rise and then start decreasing. The separatrix curve corresponds to the above quoted asymptotic value. The 10% discrepancy (or better the non negligible sensitivity upon the initial density) may be explained in terms of the systematic errors of the simulation which possibly change $A$. They are: the finite Langevin time step $\Delta\tau$, the finite time spacing $\epsilon$ and the finite spatial dimension $L$ of the lattice. Moreover, the crossover to the asymptotic regime occurs at a time which depends on the initial density.

---

[2]Let us remark that at $t = 0$ the density $\rho(t)$ is bounded to assume the fixed $\rho_0$ value. It is perfectly natural to have a decreasing intrinsic variance at small $t$.

Finally, some remarks are in order concerning the use of the Langevin algorithm in order to study criticality. In principle, one could raise the question of which is the universality class of our model at finite $\Delta\tau$. Actually, we cannot exchange the two limits $\Delta\tau \to 0$ and $T \to \infty$. The Langevin algorithm simulates exactly an action which differs from the starting one by terms proportional to powers of $\Delta\tau$. Their contribution can alter the asymptotic critical behaviour. However, at fixed $T$, we can send $\Delta\tau$ to zero and recover the correct results. After all, the spurious terms are associated to small bare couplings whose effect on the reaction may be seen only after enough time.

## V. CONCLUSIONS

In this paper we have studied the feasibility of the direct non perturbative study of a particular kind of many-body theory which can be formulated in terms of a quantum field theory. We have verified on a specific example that the method gives correct results and that it is stable, a property which was far from obvious in the interacting case. One of the important features of the method is that it can be trivially extended to more complicated processes including: higher space dimension, many species, many particle collisions; the only change is in the analytical form of the action.

## ACKNOWLEDGMENTS

We thank Ben Lee for useful discussions. We also acknowledge financial support from INFN and from the Department of Physics of the Pisa University.

# APPENDIX A: THE FEYNMAN PROPAGATOR

We use the formula

$$\int e^{-\bar{z}Az - \bar{C}z - \bar{z}B} \prod_{i=1}^{N} \frac{d^2 z_i}{\pi} = \frac{1}{\det A} \exp(\bar{C} A^{-1} B), \tag{A1}$$

and obtain for the action $S$

$$\langle z'' | e^{-t\hat{H}} | z' \rangle = \exp\left(-\frac{1}{2}|z'|^2 - \frac{1}{2}|z''|^2 + \theta^{N+1} \bar{z}'' z'\right). \tag{A2}$$

Hence, when $N \to \infty$ with $\epsilon(N+1) = t$ we get back the correct continuum result

$$\langle z'' | e^{-t\hat{H}} | z' \rangle = \exp\left(-\frac{1}{2}|z'|^2 - \frac{1}{2}|z''|^2 + e^{-t} \bar{z}'' z'\right). \tag{A3}$$

The same computation for the action $\tilde{S}$ is performed by using

$$\det A = p_N(0) = (-\alpha\beta)^{N/2}, \tag{A4}$$

$$\bar{C} A^{-1} B = (-1)^{N/2+1} \left( \beta \left(\frac{\beta}{\alpha}\right)^{N/2} \bar{z}'' z' + \alpha \left(\frac{\alpha}{\beta}\right)^{N/2} \bar{z}' z'' \right); \tag{A5}$$

and we obtain the asymptotic form when $N \to \infty$

$$U^{(N)} \sim \frac{e^t}{2^N} \exp\left\{ \frac{1}{2} e^{-t} \bar{z}'' z' - \frac{1}{2} e^t \bar{z}' z'' \right\} \to 0, \tag{A6}$$

which is not the correct result.

## 1. The Two Point Function

We study the two point function $(t_2 > t_1)$

$$G(z'', t_f | z', t_i) = \frac{1}{U(z'', t_f | z', t_i)} \langle z'' | U(t_f - t_2) \hat{a}^\dagger U(t_2 - t_1) \hat{a} U(t_1 - t_i) | z' \rangle, \quad U(t) = e^{-tH}. \tag{A7}$$

By exploiting the fact that

$$U(-t) \hat{a} U(t) = \hat{a} e^{-t}, \quad U(-t) \hat{a}^\dagger U(t) = \hat{a}^\dagger e^t, \tag{A8}$$

we obtain easily

$$G(z'', t_f | z', t_i) = e^{t_2 - t_1 - T} \bar{z}'' z', \quad T = t_f - t_i. \tag{A9}$$

Let us consider $z' = z'' = 1$, in terms of the matrix $A$ and the vectors $\bar{C}$ and $B$, the relevant expectation value is

$$\langle \bar{z}_n z_1 \rangle = (A^{-1})_{1n} + (A^{-1} B)_1 (\bar{C}^T A^{-1})_n, \quad n > 1. \tag{A10}$$

Let us begin with $S^{(N)}$, we have

$$(A^{-1} B)_1 = -\theta z', \bar{C}^T A^{-1} = \begin{pmatrix} -\theta^N \\ -\theta^{N-1} \\ \vdots \end{pmatrix} \bar{z}''. \tag{A11}$$

The correlation function is

$$\langle \bar{z}_n z_1 \rangle = \bar{z}'' z' \theta^{N-n+2} = \bar{z}'' z' \left(1 - \frac{T}{N+1}\right)^{N+2-t(N+1)/T}, n = \frac{t}{T}(N+1), \tag{A12}$$

and

$$\lim_{N \to \infty} \langle \bar{z}_n z_1 \rangle = \bar{z}'' z' e^{t-T}, \tag{A13}$$

which is the correct result. For the action $\tilde{S}$, we have explicitely

$$(A^{-1})_{1n} = \begin{cases} 0, & \text{odd } n, \\ (-1)^{n/2+1} \dfrac{\alpha^{n/2-1}}{\beta^{n/2}}, & \text{even } n, \end{cases} \tag{A14}$$

$$(A^{-1} B)_1 = (-1)^{N/2+1} \left(\frac{\alpha}{\beta}\right)^{N/2} z'', \tag{A15}$$

and

$$(\bar{C} A^{-1})_n = \begin{cases} (-1)^{N/2+1} (-1)^{(n-1)/2} \left(\dfrac{\beta}{\alpha}\right)^{(N-n+1)/2} \bar{z}'', & \text{odd } n, \\ (-1)^{n/2-1} \left(\dfrac{\alpha}{\beta}\right)^{n/2} \bar{z}', & \text{even } n. \end{cases} \tag{A16}$$

The asymptotic behaviour of the two point function $\langle \bar{z}(t)z(0)\rangle$ is then

$$|z''|^2 e^t, \tag{A17}$$

for odd $n = t/\epsilon$, and

$$e^t(z''\bar{z}'e^T - 2), \tag{A18}$$

for even $n$. In other words the continuum limit does not exist.

**Figure captions**

Figure I: Time behaviour of $\rho(t)$ starting from the initial densities: 2.0, 2.25, 2.5, 2.75, 3.0, 3.25, 3.5, 3.75, 4.0, 4.25, 4.5 and 4.75.

Figure II: Effective amplitude $\rho(t)\sqrt{t}$. Same values of the initial density as in Figure I.

Figure III: Average Im$\psi(t)$ taken from the run at $\rho(0) = 2.0$.

Figure IV: Time history of $\rho(100)$ taken from the run at $\rho(0) = 2.0$.